\documentclass[12pt,preprint]{aastex}

\newcommand{\msun} {$M_{\sun}$}

\newcommand{\Te} {T_{\rm eff}}
 
\begin{document}

\title{The Ratio of Helium- to Hydrogen-Atmosphere White Dwarfs: 
Direct Evidence for Convective Mixing}

\author{P.-E. Tremblay and P. Bergeron}
\affil{D\'epartement de Physique, Universit\'e de Montr\'eal, C.P.~6128, 
Succ.~Centre-Ville, Montr\'eal, Qu\'ebec, Canada, H3C 3J7.}
\email{tremblay@astro.umontreal.ca, bergeron@astro.umontreal.ca}

\begin{abstract}

We determine the ratio of helium- to hydrogen-atmosphere white dwarf
stars as a function of effective temperature from a model atmosphere
analysis of the infrared photometric data from the Two Micron All Sky
Survey combined with available visual magnitudes. Our study
surpasses any previous analysis of this kind both in terms of the
accuracy of the $\Te$ determinations as well as the size of the
sample. We observe that the ratio of helium- to hydrogen-atmosphere
white dwarfs increases gradually from a constant value of $\sim0.25$
between $\Te=15,000$~K and 10,000~K to a value twice as large in the
range $10,000 > \Te > 8000$ K, suggesting that convective mixing,
which occurs when the bottom of the hydrogen convection zone reaches
the underlying convective helium envelope, is responsible for this
gradual transition. The comparison of our results with an approximate
model used to describe the outcome of this convective mixing process
implies hydrogen mass layers in the range $M_{\rm H}/M_{\rm
tot}=10^{-10}$ to $10^{-8}$ for about 15\% of the DA stars that 
survived the DA to DB transition near $\Te\sim 30,000$~K, the remainder
having presumably more massive layers above $M_{\rm H}/M_{\rm
tot}\sim10^{-6}$.

\end{abstract}


\section{INTRODUCTION}

The Two Micron All Sky Survey (2MASS) represents one the largest set
of homogeneous infrared photometric data for all areas of the sky
applicable to white dwarf stars of various spectral types. Indeed, the
2MASS survey contains $JHK_S$ magnitudes for almost all point sources
of the sky up to the magnitude thresholds of $J \sim 16.8$, $H \sim
16.5$, and $K_S\sim 16.2$ for detections with any formal
uncertainty \citep[see for instance Fig.~1 of][]{tremblay07}. This yields a nearly
complete magnitude limited survey of white dwarfs that allows us to
trace a portrait of the local population of white dwarfs, assuming we
can properly identify the white dwarfs in this survey. To that effect,
the most complete compilation is that of the Villanova White Dwarf
Catalog\footnote{http://www.astronomy.villanova.edu/WDCatalog/index.html}
(WDC), which contains more than 5500 spectroscopically identified
white dwarfs reported in the literature. The interest of such a
sizeable sample of homogeneous infrared photometric observations is to
reach an unprecedented level in the statistical analysis of white
dwarf stars. In addition, \citet{tremblay07} have demonstrated the
overall reliability of the 2MASS Point-Source Catalog (PSC) photometry
for the purpose of detailed comparisons with the predictions from
white dwarf model atmospheres.

Studies of the spectral evolution of white dwarfs attempt to
understand how the surface composition and the corresponding spectral
signature evolve along the white dwarf cooling sequence (see
\citealt{fontaine87,fontaine91} for a review). Because of the intense 
gravitational field present in these stars, hydrogen will float to the
surface in the absence of competing mechanisms, while heavier elements
will sink below the photospheric regions.  Indeed, the majority of
white dwarfs have optical spectra that are completely dominated by
strong hydrogen Balmer lines --- the DA stars --- with
hydrogen-dominated atmospheres. We also find that almost 25\% of white
dwarfs have helium-dominated atmospheres, mostly DO, DB, DQ, and DZ
stars, and some of the DC stars as well. The fact that not all white
dwarfs have hydrogen-rich atmospheres imply that either some stars are
born with hydrogen-deficient atmospheres and remain that way
throughout their evolution, or alternatively, that several physical
mechanisms such as diffusion, accretion, convection, radiation
pressure, and stellar winds are competing with gravitational settling
in determining their surface composition as they cool off. This
particular issue can be resolved by analyzing the distribution of
spectral types as a function of effective temperature (or any
temperature index) for a large sample of white dwarf stars, in a way
similar to the analysis of \citet{sion84} nearly twenty five years
ago.

Before discussing the physical processes that will affect the spectral
evolution of hydrogen- and helium-atmosphere white dwarfs, it is
important first to consider the constraints brought about by the
studies of the hottest white dwarfs. The canonical mass fractions of
light elements in white dwarfs considering the mass threshold for
residual nuclear burning are $M_{\rm He}/M_{\rm tot} \sim 10^{-2}$ and
$M_{\rm H}/M_{\rm tot} \sim 10^{-4}$. However, multiple excursions on
the AGB nuclear burning phase, during which very late helium flashes
could remove essentially all of the hydrogen \citep{werner06}, could also produce
white dwarfs with total hydrogen masses much lower than the canonical
value of $10^{-4}$. The first class would correspond to the
progenitors of the hottest DA stars while the second class would
represent the progenitors of hot white dwarf stars with
hydrogen-deficient atmospheres --- the PG 1159 stars at very high
effective temperatures, and the DO stars at slightly lower
temperatures, whose spectra are dominated by He\textsc{ii} lines. It
has traditionally been believed that by the time DO stars cool off to
$\Te\sim45,000$~K, residual amounts of hydrogen ($M_{\rm H}/M_{\rm
tot} \sim 10^{-16}$) thoroughly mixed in their helium-rich envelope
would gradually accumulate to the surface, turning all
helium-atmosphere white dwarfs into DA stars
\citep{fontaine87,fontaine91}. This scenario would account
for the existence of the so-called DB gap, a range of temperature
between $\Te=45,000$~K and 30,000~K where all white dwarfs have a DA spectral
type. However, the recent discovery in the Sloan Digital Sky Survey
of several DB stars within the gap \citep{eisenstein06a} implies that the total mass
of hydrogen left in the envelope of DO stars can be even smaller than
previously believed. It is then generally admitted that the
significant increase in the number of helium-atmosphere DB white
dwarfs below $\Te\sim30,000$~K can be explained in terms of the convective
dilution of the superficial hydrogen atmosphere by the underlying
helium convective envelope, provided that the hydrogen layer is
sufficiently thin
\citep[$M_{\rm H}/M_{\rm tot} \sim 10^{-15}$,][]{macdonald91}.

At even lower effective temperatures ($\Te\lesssim12,000$~K),
hydrogen-atmosphere white dwarfs get a second opportunity to evolve
into stars with helium-dominated atmospheres when the superficial
hydrogen layer becomes convective over a significant fraction of its
depth. This process is illustrated in Figure \ref{fg:f1} where the
extent of the hydrogen convection zone is displayed as a function of
decreasing effective temperature for a 0.6 \msun\ DA white dwarf,
based on evolutionary models with thick hydrogen layers similar to
those described by \citet{fon01} and kindly provided to us by
G.~Fontaine and P.~Brassard. These calculations show that if the
hydrogen envelope is thin enough, the bottom of the hydrogen
convection zone may eventually reach the underlying and more massive
convective helium layer, resulting in a mixing of the hydrogen and
helium layers
\citep{strittmatter71,shipman72,baglin73,koester76,vauclair77}. Figure \ref{fg:f1}
also indicates that the effective temperature at which this mixing
occurs will depend on the thickness of the hydrogen envelope. The
thicker the envelope, the lower the mixing temperature; if the
hydrogen layer is more massive than $M_{\rm H}/M_{\rm tot} \sim
10^{-6}$, mixing will never occur.

The simplest physical model that can be used to describe this
convective mixing process is to assume that hydrogen and helium are
homogeneously mixed. Since the helium convection zone is much more
massive ($M_{\rm He-conv}/M_{\rm tot}\sim10^{-6}$) than the hydrogen
layer when mixing occurs, it is generally assumed that a DA star would
be transformed into a helium-atmosphere white dwarf of type DB, DQ,
DZ, or DC with only a trace abundance of hydrogen. More detailed
calculations discussed by \cite{fontaine91} confirm these
predictions. If such a process takes place in cool white dwarfs, we
then expect the ratio of helium- to hydrogen-atmosphere white dwarfs
to increase at low effective temperatures.  A comparison between the
observed ratio as a function of $\Te$ and the theoretical expectations
would make it possible to estimate the thickness of the hydrogen
layers in DA white dwarfs. In turn, these determinations could be
compared to independent measurements of the hydrogen layer mass
inferred from ZZ Ceti asteroseismology.

The best approach to study the occurrence of this convective mixing
process begins with the statistical analysis of large samples of cool
white dwarfs for which we can determine the main atmospheric
constituent and the effective temperature, or any analogue temperature
index such as the absolute $V$ magnitude. \citet{sion84} was the first
to study this problem by estimating the non-DA to DA ratio at low
effective temperatures using a proper motion sample of 695 white
dwarfs drawn from the catalog of spectroscopically identified white
dwarfs available at that time. The results from Figure 1 of
\cite{sion84} are reproduced here in Figure \ref{fg:f2} in terms of
the non-DA to DA ratio (rather than the absolute number of DA and
non-DA stars) as a function of the absolute visual magnitude $M_V$.
Looking at the results, we notice a first increase in this ratio at
$M_V>11.25$, which corresponds to $\Te\sim15,000$ K for a 0.6 \msun\
white dwarf, a temperature much in excess of the theoretically
predicted value for the convective mixing process.  A second increase
occurs at $M_V\sim 12.5$, or $\Te \sim 10000$ K, but the observed
ratio drops suddenly afterwards and this increase is probably not
significant. Then a third increase occurs at $M_V> 13$, or
$\Te<8000$~K.  Clearly, the evidence based on these results are not
exactly convincing, and our current picture of the situation below
12,000 K is at best sketchy.  The global portrait did not change
significantly since the analysis of
\citet{sion84}, and this remains the only available proof quoted in the literature
of the transformation of some DA stars into non-DA stars
(see also \citealt{greenstein86}).

Our goal is to improve upon the analysis of \citet{sion84} by using
the 2MASS photometric sample combined with the WDC database to
identify white dwarfs and determine their atmospheric composition, and
most importantly to obtain more accurate temperature determinations
using a full model atmosphere analysis of the $VJHK_S$ photometry. We
first describe in \S~2 the 2MASS photometric white dwarf sample used
in our analysis. The determination of the atmospheric parameters is
then presented in \S~3 and the uncertainties related to our approach
are discussed at length in \S~4. The evolution of the ratio of helium-
to hydrogen-atmosphere white dwarfs as a function of effective
temperature is then determined in \S~5. Our conclusions follow in
\S~6.

\section{DEFINITION OF THE 2MASS WHITE DWARF SAMPLE}

Our starting point is the Villanova White Dwarf Catalog (WDC) of
spectroscopically identified white dwarfs.  We took the most recent
electronic version of the catalog (August 2006), which includes a
total of 5557 white dwarfs. We made a first selection by taking only
the objects with at least one published visual magnitude\footnote{We
have neglected magnitudes prior to 1972, photographic magnitudes, $B$
magnitudes, as well as those flagged as uncertain.} and with at least
one measurement brighter than $V=18.5$. Figure
\ref{fg:f3} shows the predicted $V$ magnitude for which $H_{\rm 2MASS}$ = 16.5 --- a
representative value for the 2MASS detection limit in this band --- as
a function of $\Te$ for our pure hydrogen and pure helium model
atmospheres at $\log g=8$. Only white dwarfs located below these
curves at a given effective temperature can be detected by 2MASS in
the $J$ and $H$ bandpasses. For white dwarfs with $\Te >$ 5000 K, the
$V < 18.5$ criterion is sufficient to find a maximum number of white
dwarfs. Finally, since our study is aimed at tracing the spectral
evolution of cool white dwarfs, we neglected all objects with early
spectral classifications (e.g., PG 1159, DO, etc.) or for which the
spectral classification was unavailable altogether. Thus, a total of
1473 remaining white dwarf stars define our initial WDC sample.

We then searched the 2MASS PSC for $JHK_S$ photometric observations of
all white dwarfs from this initial WDC sample. We used the GATOR batch
file tool and a $20''$ search window centered on a set of improved
coordinates measured by J.~B.~Holberg (2005, private
communication). In most instances, multiple sources were found within
the search window and we unambiguously identified each object by
comparing the 2MASS atlas with the finding charts available from the
online version of the Villanova WDC. For several objects close to the
detection limits, only one or two magnitudes are detected with formal
uncertainties. We neglect all objects with only one detected magnitude
since in such cases, only one degree of freedom would be available in
our minimization procedure described in \S~3.  We recovered 940
objects in the 2MASS PSC from the 1473 objects contained in our
initial WDC sample. The remaining 533 objects are eliminated for the
following reasons: 508 are too faint to be detected by 2MASS in at
least two bands, 15 are bright enough in the 2MASS atlas but are not
part of the 2MASS PSC, and 10 could not be unambiguously identified
from the comparison of the 2MASS atlas and the published finding
charts.

A spectral energy distribution is then built for each object by
combining the 2MASS $JHK_S$ photometry with the $V$ magnitude (or
Str\"omgren $y$) taken from the WDC. In addition, about 10\% of our
$V$ magnitudes come from unpublished USNO photometry
\citep{dahn07}. When more than one visual magnitude is available, we
take an unweighted average, unless we had reasons to believe that a
given measurement was less accurate than others. We also assume a
typical uncertainty of 0.05 mag for the visual magnitudes.

\section{ATMOSPHERIC PARAMETER DETERMINATIONS}

We classify the white dwarfs in our sample into two broad categories,
those with hydrogen atmospheres (DA stars, including DAB and DAZ
stars) and those with helium atmospheres (DB, DZ, and DQ stars as well
as helium-atmosphere objects with weak hydrogen features such as
Ross 640 and L745-46A; the case of DC stars is discussed further
below). When the spectral classification is ambiguous, we check in the
literature to confirm the presence of hydrogen lines compatible with
the estimated effective temperature. We then fit the $VJHK_S$ (or
$VJH$) photometry with our grids of pure hydrogen or pure helium model
atmospheres described in
\citet{tremblay07} and references therein.  Our fitting technique is 
similar to that outlined in \citet{bergeron97,bergeron01}. Briefly,
the magnitudes are converted into monochromatic fluxes using the zero
points defined in \citet{holberg06} for photon counting devices, but
using the transmission function of \citet{bessell90} for $V$ and the
2MASS transmission functions of \citet{cohen03} for $JHK_S$. The
resulting fluxes are then compared with those predicted from the model
atmospheres, properly averaged over the same bandpasses.  The
effective temperature and the solid angle $\pi(R/D)^2$ (with $R$ the
radius of the star and $D$ the distance to Earth) are considered free
parameters in the $\chi ^2$ minimization procedure. We assume a value
of $\log g=8.0$ throughout.  The surface gravity has very little
effect on the predicted $(V-J)$ color indices of white dwarfs. The
results shown in Figure \ref{fg:f4} indicate that a difference of 1
dex in $\log g$ corresponds to a color difference smaller than 0.05
mag at all temperatures. This slight difference, which is of the order
of our color uncertainties, allows us to fix the surface gravity at
$\log g=8.0$ for all objects. Since the mass distributions of
hydrogen- and helium-atmosphere white dwarfs have to a good
approximation the same mean value
\citep{bergeron92,beauchamp95,liebert05,voss07}, this simplifying
assumption should not affect the results of our analysis
significantly.

For the featureless DC stars, we consider both hydrogen- and helium-atmosphere
solutions.  If the hydrogen solution is above $\Te=5000$~K, we adopt
the helium solution since in principle, H$\alpha$ should have been
detected spectroscopically according to the photometric analyses of
\citet{bergeron97,bergeron01}. However, if the hydrogen solution is
below this temperature, it is impossible to determine the main
atmospheric constituent from the restricted available photometry, and
therefore we set the limit of our analysis to $\Te>5000$~K.

There is also a significant fraction of binary white dwarfs in our
sample. If the companion is a cool main-sequence star, the observed
$JHK_S$ fluxes are substantially larger than those predicted by the
models, and the $(J-H)$ color index takes a large positive value
\citep{tremblay07}, which typically yields a bad fit to the energy
distribution. We eliminated 90 binary candidates satisfying at least
two of the following criteria\footnote{16 of these did not exactly
meet two criteria but they have poor fits and atmospheric parameters
that are incompatible with the published spectral type.}: (1) a
photometric temperature below 6000 K incompatible with the spectral
type, (2) an observed flux at $H$ larger than that at $J$, and (3) a
flux at $J$ larger than ten times the flux at $V$. We have also
eliminated 9 binary systems that are partially resolved in the 2MASS
atlas, but whose colors in the the 2MASS PSC are likely to be
contaminated. The large majority of our binary candidates have been
previously identified as such in the literature.

Sample fits for eight bright ($11 < V < 13.5$) and well studied white
dwarfs in our sample are displayed in Figure \ref{fg:f5}. We notice in
each case that the best fit is in excellent agreement with the
$VJHK_S$ photometry, for both hydrogen- or helium-atmosphere white
dwarfs. We also compare in Table 1 the photometric temperatures for
these objects with independent spectroscopic or photometric effective
temperature determinations reported in the literature. This comparison
indicates that our technique applied to cool white dwarfs of various
spectral types yields $\Te$ values with a precision of the order of
5\% compared to other methods, provided we have accurate photometric
data.

\section{ATMOSPHERIC PARAMETER UNCERTAINTIES}

We attempt in this section to better quantify the uncertainties of our
effective temperature determinations in order to determine the optimal
bin widths for the histograms used in our study of convective mixing,
presented in section \S~5. These uncertainties can be first quantified
by looking at the rate of change of the $(V-J)$ color index as a
function of $\Te$. We show in Figure \ref{fg:f6} the effective
temperature variation, $\Delta \Te$, for a corresponding change of the
$(V-J)$ color index of $\Delta (V-J)=0.08$, a value that defines the
mean uncertainty of this color index in our sample. We notice that the
$\Te$ uncertainties increase rapidly towards high effective
temperatures, and for this reason, we restrict our analysis to
$\Te<15,000$~K.

We compare in Figure \ref{fg:f7} our photometric $\Te$ determinations
for 133 cool white dwarfs in common with the analyses of
\citet{bergeron97,bergeron01}, which are based on independent
$BVRIJHK$ photometric data and in several cases, trigonometric
parallax measurements from which surface gravities can be
determined. To be internally consistent, we have revised their
atmospheric parameters using our own fitting procedure based on
the slightly different synthetic photometric calibration of \citet{holberg06}.
As well, we consider only the stars that have been observed in at
least the $J$ and $H$ bands. We can see that the agreement is
excellent in this temperature range and that the $VJHK_S$ photometry
is sufficient to constrain the $\Te$ values with the precision sought
in our analysis.

We present in Figure \ref{fg:f8} a similar comparison of our
photometric temperatures with independent values obtained from line
profile fitting of DA and DB stars. The spectroscopic
temperatures for the 198 DA stars in common with our sample are taken
from the ongoing spectroscopic survey of A.~Gianninas (2007, private
communication). We notice in this case that both samples are generally
in good agreement, but several discrepancies are also observed where
the spectroscopic temperatures are significantly larger than the
photometric values ($\Delta \Te>0$), in particular at high
temperatures. A first explanation for this discrepancy is that above
$\Te > 11,000$~ K, the $(V-J)$ color index is a slowly varying
function of effective temperature (see Fig. \ref{fg:f6}), and thus
white dwarfs with large temperature uncertainties contaminate this
particular region. This is confirmed by the fact that most white
dwarfs showing a large discrepancy are near the 2MASS detection limit
(objects not detected at $K_S$ in 2MASS are shown by open circles in
Fig. \ref{fg:f8}). Therefore, we add the constraint that only white
dwarfs detected in all three $JHK_S$ bands are allowed in our analysis
for $\Te>11,000$~K. The second group of objects showing a positive
$\Delta \Te$ are unresolved double degenerate systems, three of which
are identified in Figure \ref{fg:f8} by filled squares. In all cases
the photometric temperature is substantially lower than the
spectroscopic value, and both spectroscopic and photometric fits are
excellent. It is therefore nearly impossible to eliminate double white
dwarf systems from our analysis. There remain 10 other objects with a
$\Delta \Te$ larger than 20\% for which the photometric fit is good
and there is no clear explanation for a discrepancy. These could
also correspond to unresolved degenerate binaries, although a
closer inspection rather suggests that the $V$ magnitudes, which are
old measurements, are probably in error.  

In the bottom panel of Figure \ref{fg:f8} we compare our photometric
temperatures for 18 DB stars with those determined spectroscopically
by \citet{beauchamp95,beauchamp96}. We see that the agreement is excellent
although the spectroscopic temperatures are systematically larger than
the photometric values by about 5\%. This small shift could be due to
the neglect of small traces of hydrogen or metals in both analyses, or
to inaccuracies in the treatment of the broadening theory for neutral
helium lines at low effective temperatures, which will mainly affect
the spectroscopic temperatures \citep[see, e.g.,][]{kepler07}.  The
effective temperature dispersions observed in the figures above
indicate that histograms with bin widths of 1000 K below
$\Te=11,000$~K and of 2000~K above this temperature are
acceptable. These values are reproduced in Figures \ref{fg:f6} to
\ref{fg:f8}.

\section{RESULTS ON CONVECTIVE MIXING}

We present in Table 2 the number of hydrogen- and helium-atmosphere
white dwarfs in different effective temperature bins for our final
2MASS sample of 447 objects (after the temperature cutoff at
$\Te=15,000$ and $5000$~K and the removal of 2MASS data with low
accuracy above $\Te=11,000$~K have been applied). This sample is nearly
twice as large as that used in the analysis of
\citet{sion84} in the same range of effective temperature.
We also display in Figure \ref{fg:f9} the corresponding histogram of
the number of hydrogen- and helium-atmosphere white dwarfs as a
function of $\Te$.  The total number of stars in
each bin below $\Te=11,000$~K is comparable, a result that can be
explained by the competition between the increasing luminosity
function and the decreasing detection probability of cooler, fainter
objects. We further observe that hydrogen-atmosphere DA white dwarfs
are always the dominant type. Consequently, convective mixing --- if
this process occurs at all --- should affect only a small fraction of
cool DA stars.

For a magnitude limited sample --- here the $H$ magnitude ---, we
probe a limited volume of space. This volume is defined in terms of
the distance at which a given white dwarf can be detected at the
limiting magnitude of the survey. However, this distance is different
for hydrogen- and helium-atmosphere white dwarfs because they have
different continuum opacity sources and corresponding monochromatic
fluxes. Figure \ref{fg:f10} presents the ratio of the volume probed by
hydrogen and helium white dwarfs as a function of effective
temperature for a magnitude limit in the $H$ bandpass. While the
assumption of a fixed value of $\log g$ was appropriate for
determining the effective temperatures, this may not be the case for
the volume effect discussed here. Indeed, white dwarfs stars with
large radii (or low masses) can be detected at larger distances. The
differences in the mass distributions of DA and DB stars could
therefore affect the ratio \citep[see also][]{eisenstein06a}. While
the results of \cite{beauchamp96} suggest that the mass distribution
of DB stars is narrower than that of DA stars, the recent results of
\citet{voss07} reveal instead that both distributions are comparable.
Hence the small residual differences in the low and high mass tails
are not expected to affect our results significantly.

To take into account the volume effect, we henceforth correct the
ratio of the number of helium- to hydrogen-atmosphere white dwarfs,
$N_{\rm He-atm}/N_{\rm H-atm}$, by the corresponding factor given in
Figure 11. This ratio corrected for the volume effect is reported in
Table 2 and displayed in Figure \ref{fg:f11} as a function of
effective temperature. The error bars are statistical and they are
mainly governed by the limited number of helium-atmosphere objects in
each bin. We also show in the same figure the ratio obtained by
shifting the temperature bins in Figure \ref{fg:f9} by 500 K.

We can immediately notice one important trend in the observed ratio of
helium- to hydrogen-atmosphere white dwarfs displayed in Figure
\ref{fg:f11}.  At high effective temperatures, this ratio takes a
constant value of $\sim0.25$ and it shows a gradual increase from
$\Te\sim10,000$~K to a fairly constant value twice as large
($\sim0.5$) at 8000~K. The only physical mechanism that could account
for this increase is the convective mixing of thin hydrogen layer with
the more massive underlying helium envelope. These results further
imply that about 15\% of all DA stars\footnote{The transformation
implies the following change of ratio $N_{\rm He-atm}/N_{\rm
H-atm}$~$\rightarrow$~$(N_{\rm He-atm}+N_{\rm H-transformed})/(N_{\rm
H-atm}-N_{\rm H-transformed})$.} between $\Te=15,000$~K and 10,000~K
have turned into helium-atmosphere white dwarfs --- either DQ, DZ (or
DZA), or DC --- by the time they reach 8000~K. Note that the
results of our analysis become too uncertain above the temperature
range displayed here (see Fig.~\ref{fg:f6}).

We can gain a better understanding of our results by looking at the
simple theoretical model of convective mixing discussed in the
Introduction. As the white dwarf cools off, the bottom of the hydrogen
convection zone will eventually reach the helium convection zone,
provided the hydrogen layer is thin enough (see Fig.~\ref{fg:f1}),
with the mixing temperature depending on the thickness of the hydrogen
layer. White dwarfs with thicker hydrogen layers will mix at lower
effective temperatures.  A comparison of the results shown in Figures
\ref{fg:f1} and \ref{fg:f11} implies that there are very few DA stars
between $\Te\sim15,000$~K and 10,000~K --- the region where ZZ Ceti
stars are found --- with hydrogen layer masses smaller than $M_{\rm
H}/M_{\rm tot}\sim10^{-10}$. The gradual increase of the observed
ratio from 10,000~K down to 8000~K further implies that only $15\%$ of
DA white dwarfs in the 10$-$15,000~K temperature range have
superficial hydrogen layers with a mass in the range $M_{\rm H}/M_{\rm
tot}\sim10^{-10}$ to $10^{-8}$. Note that this range of layer mass
depends sensitively on the assumed convective efficiency
(ML2/$\alpha=0.6$ in Fig.~\ref{fg:f1}). Envelope models with a more
efficient parameterization of the convective energy transport would
imply even thicker hydrogen layers \citep[see Fig.~9
of][]{tassoul90}. The lack of evidence for convective mixing at lower
effective temperatures between $\Te=8000$~K and 5000~K implies a
discontinuity in the hydrogen layer mass distribution, with the large
majority ($\sim85$\%) of all DA stars between $\Te=15,000$~K and
10,000~K having hydrogen layer masses above $M_{\rm H}/M_{\rm
tot}\sim10^{-6}$ or so. The constant ratio of $\sim 0.25$ in this
temperature range also implies that $\sim 20$\% of all DA stars in the
DB gap (or DB ``trough'') between $\Te=45,000$~K and 30,000~K have
turned into DB stars by the time they reach 15,000~K; we cannot tell
from our results at which effective temperature the DA to DB
transition occurs precisely.

The numbers given above could change slightly if one takes into
account the presence of traces of metals (including carbon) and
hydrogen in the model atmosphere calculations of DQ and DZ
stars. Indeed, the atmospheric structure of cool helium-atmosphere
white dwarfs is affected significantly by small traces of metals or
hydrogen, which then provide most of the free electrons
\citep{provencal02, dufour05, dufour07}. These additional electrons
increase the contribution of the He$^-$ free-free opacity, which in
turn reduces the atmospheric pressure.  The results of Dufour et
al.~(2005, 2007) reveal that photometric temperatures of DQ and DZ
stars can be systematically reduced by 500~K on average when metals
are included in the model calculations.  Therefore, this could affect
the precise temperature dependence of the change in the ratio of
helium- to hydrogen-atmosphere white dwarfs observed in Figure
\ref{fg:f11}. 

We finally note that our results are compatible with the
asteroseismological analyses of ZZ Ceti stars, which suggest that most
ZZ Ceti stars have fairly thick hydrogen layers of the order of
$M_{\rm H}/M_{\rm tot}\sim 10^{-6}$ and higher
\citep[see, e.g.,][]{fontaine92,bergeron93,bradley98,brassard06,pech06}. 
Only a small fraction (about 15\%) of all ZZ Ceti stars should have
thin layers according to our results. Furthermore, our results are
also in qualitative agreement with the analysis of DZ stars by
\citet{dufour07} who suggested that the presence of hydrogen in cool
DZA white dwarfs could be explained better as the outcome of
convectively mixed DA stars rather than due to accretion from the
interstellar medium. This suggestion is based on estimates of the
total mass of hydrogen present in the mixed hydrogen and helium
envelope of DZ and DZA stars. Measurements of the hydrogen abundances
in these stars suggest hydrogen layer masses in about the same range
as that inferred here
\citep[see Fig.~13 of][]{dufour07}.

\section{CONCLUSION}

We have presented a detailed study of the spectral evolution of cool
white dwarfs following the pioneering effort of \citet{sion84}. In
particular, we have determined the ratio of helium- to
hydrogen-atmosphere white dwarfs below $\Te=15,000$~K as a function of
effective temperature by increasing the size of the sample by a factor
of 2, and by obtaining more precise measurements of the effective
temperature of these stars through model atmosphere fits to Johnson
$V$ (or Str\"omgren $y$) and 2MASS $JHK_S$ photometric
observations. We have confirmed that in this temperature range, the
evolution of hydrogen- and helium-atmosphere white dwarfs is not
independent, and we have revealed in greater detail the role of the
convective mixing process responsible for the coupling between these
two white dwarf types. We have found that about 15$\%$ of cool
hydrogen-atmosphere DA white dwarfs between $\Te=15,000$~K and
10,000~K are transformed into helium-atmosphere non-DA white dwarfs at
lower temperatures. Using a basic model for convective mixing, our
results imply in turn that 15\% of the DA stars that have survived the
DA to DB transition have hydrogen layer masses in the range $M_{\rm
H}/M_{\rm tot}\sim10^{-10}$ to $10^{-8}$, although the exact values
depend on the assumed convective efficiency in the evolutionary model
calculations. The remaining DA stars would presumably have more
massive hydrogen layers, probably in excess of $M_{\rm H}/M_{\rm
tot}\sim10^{-6}$.

Further analysis of the convective mixing problem would require a
significantly larger photometric sample, such as the {\it ugriz}
photometric sample of the Sloan Digital Sky Survey (SDSS). This sample
covers a few areas of the sky with a completeness of around 95\% up to
$g=20$. However, most spectroscopic follow-ups have relied on
color-color diagrams to identify white dwarfs
\citep{kl04,eisenstein06b}, and this method is strongly biased towards
hot white dwarfs ($\Te > 12000$~K). Therefore, the SDSS white dwarfs
identified so far cannot be used to study the convective mixing
process at low effective temperatures. One way to proceed to identify
cooler white dwarfs in the SDSS is through proper motion-color
diagrams. \citet{harris06} have used this method to identify nearly
6000 white dwarfs but a complete spectroscopic analysis similar to
that of \citet{kilic06} has yet to be performed to take full advantage
of this sample.

\acknowledgements
We thank A.~Gianninas for a careful reading of our manuscript, and
G.~Fontaine and P.~Brassard for providing with us with the details of
their evolutionary models. We are also grateful to H.~C.~Harris for
allowing us to use some of the USNO $V$ magnitudes prior to
publication. This work was supported in part by the NSERC Canada and
by the Fund FQRNT (Qu\'ebec). P. Bergeron is a Cottrell Scholar of
Research Corporation.

\clearpage
\clearpage
\begin{deluxetable}{lcrrc}
\tabletypesize{\scriptsize}
\tablecolumns{5}
\tablewidth{0pt}
\tablecaption{Comparison of Effective Temperature Determinations}
\tablehead{
\colhead{} &
\colhead{} &
\colhead{$T_{\rm eff}$ (K)} &
\colhead{$T_{\rm eff}$ (K)} &
\colhead{} \\
\colhead{WD} &
\colhead{Spectral Type} &
\colhead{This study} &
\colhead{Others} &
\colhead{Reference} 
}
\startdata
0046$+$051  &  DZ   &  6716  & 6220  & 2\\
0135$-$052  &  DA   &  7353  & 7275   & 1\\
0310$-$688  &  DA   &  15891 & 15500  & 1\\
1142$-$685  &  DQ   &  8429  & 7900  & 3\\
1647$+$591  &  DAV  &  12557 & 12258  & 1\\ 
1917$+$077  &  DBQA &  11488 & 10200       & 4\\ 
1940$+$374  &  DB   &  15565 & 16877       & 5\\ 
2032$+$248  &  DA   &  20287 & 19511 & 1\\
\enddata
\tablecomments{
References: (1) A.~Gianninas (2007, private communication), (2) \citet{dufour07}, (3) \citet{dufour05}, (4) \citet{oswalt91}, (5) \citet{beauchamp96}.}
\end{deluxetable}

\clearpage
\clearpage
\begin{deluxetable}{cccc}
\tabletypesize{\scriptsize}
\tablecolumns{4}
\tablewidth{0pt}
\tablecaption{Ratio of He- to H-atmosphere white dwarfs}
\tablehead{
\colhead{$T_{\rm eff}$ range} &
\colhead{} &
\colhead{} &
\colhead{} \\
\colhead{(10$^{3}$ K)} &
\colhead{$N_{\rm He-atm}$} &
\colhead{$N_{\rm H-atm}$} &
\colhead{Ratio\tablenotemark{a}} 
}
\startdata
15$-$13  &  11   &  41 & 0.268  \\
13$-$11  &   9   &  38 & 0.259  \\
11$-$10  &  10   &  44 & 0.265  \\
10$-$9  &  12   &  41 &  0.347  \\
9$-$8  &  14   &  38 &  0.444 \\
8$-$7  &  19   &  49 &  0.485 \\
7$-$6  &  18   &  51 &  0.454 \\
6$-$5  &  14   &  38 &  0.448 \\
\enddata
\tablenotetext{a}{The ratio is corrected for the volume effect discussed in the text.}
\end{deluxetable}

\clearpage

\figcaption[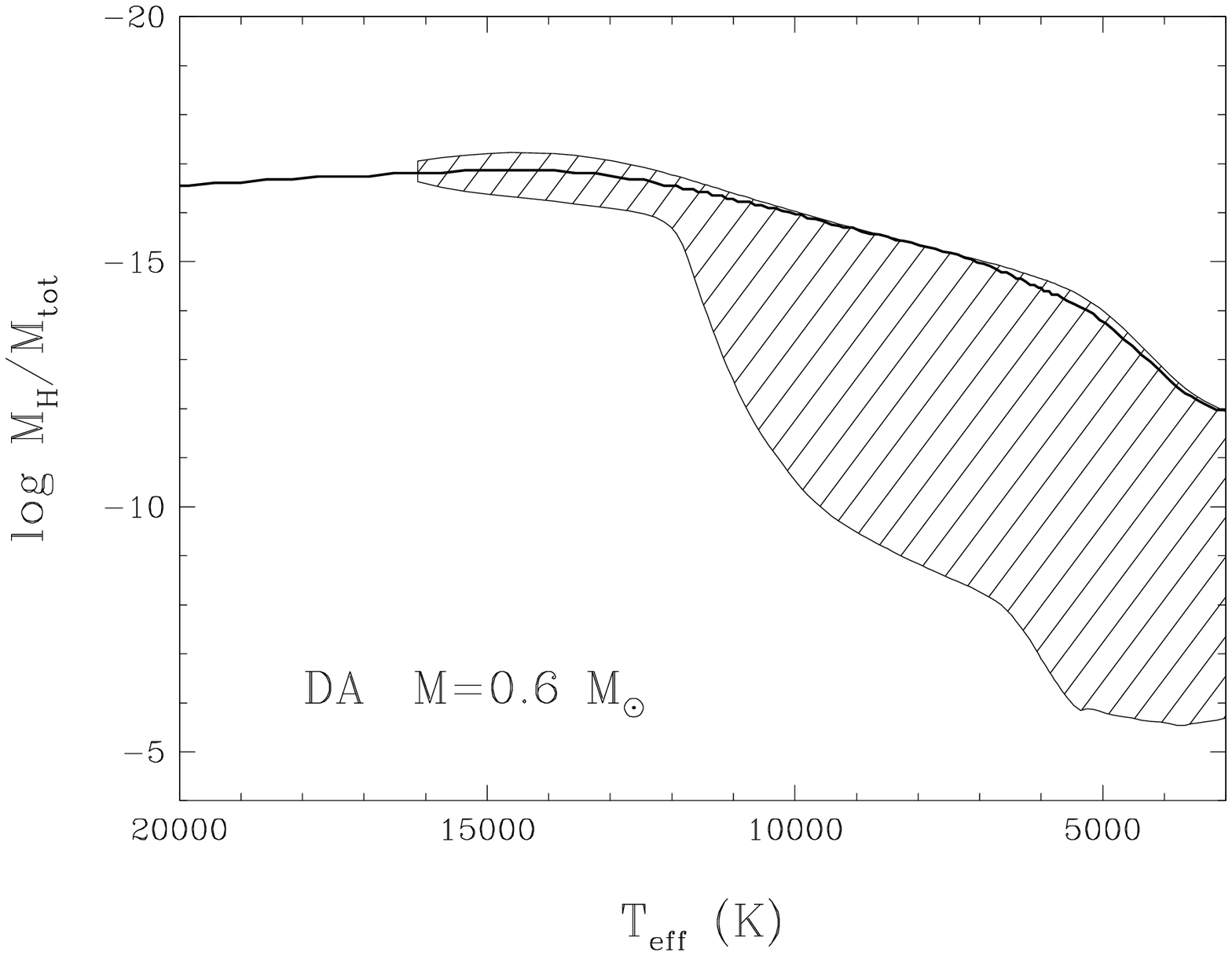] 
{Location of the hydrogen convection zone ({\it hatched region}) as a
function of effective temperature in the pure hydrogen envelope of a 0.6
\msun\ DA white dwarf calculated with the ML2/$\alpha=0.6$ version of the
mixing-length theory (from G.~Fontaine \& P.~Brassard 2006, private
communication). The depth is expressed as the fractional mass above
the point of interest with respect to the total mass of the star. The
thick solid line corresponds to the photosphere ($\tau_R\sim1$).
\label{fg:f1}}

\figcaption[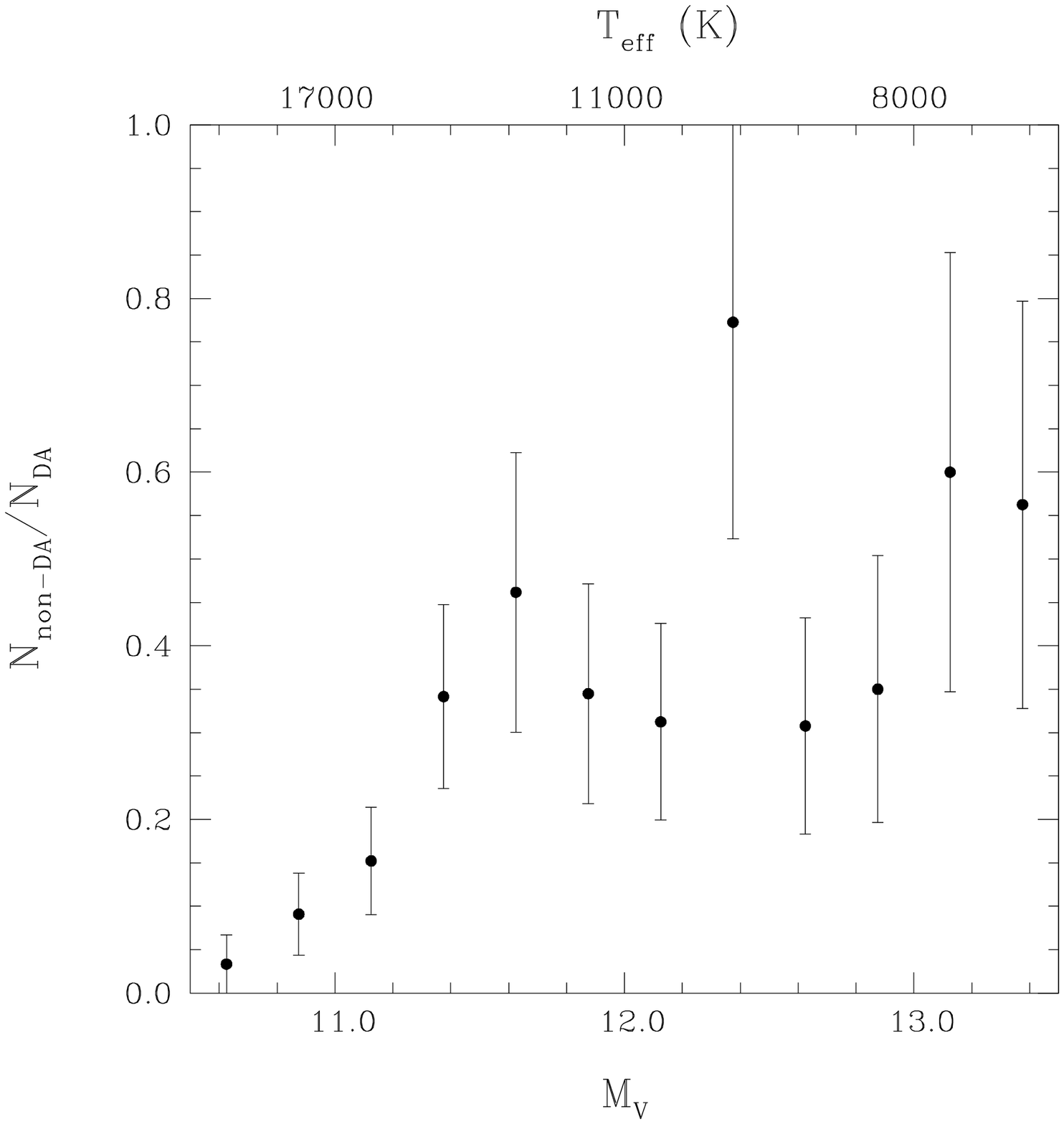] 
{The ratio of non-DA to DA white dwarfs as a function of absolute
visual magnitude for the sample of \citet{sion84}. The error bars
correspond to statistical uncertainties. The temperature scale for
pure hydrogen model atmospheres with $M=0.6$ \msun\ is shown for
comparison at the top of the figure.
\label{fg:f2}}

\figcaption[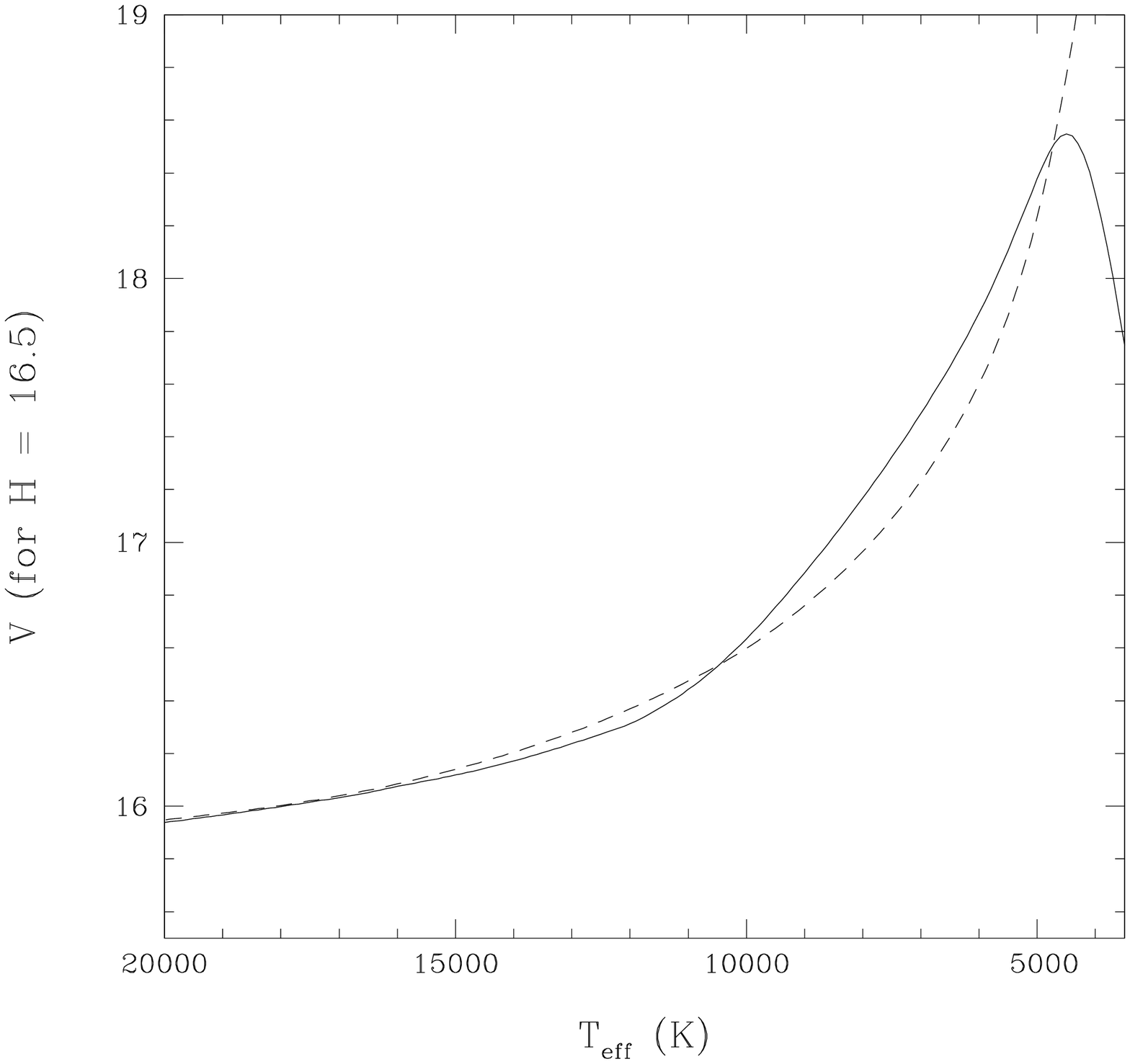]
{Predicted $V$ magnitude as a function of $\Te$ for pure hydrogen
({\it solid line}) and pure helium ({\it dashed line}) model
atmospheres at $\log g=8$ and for which $H_{\rm 2MASS}=16.5$. The
turnover of the hydrogen-atmosphere curve below 5000 K is due to the
onset of the collision-induced absorptions by molecular hydrogen.
\label{fg:f3}}

\figcaption[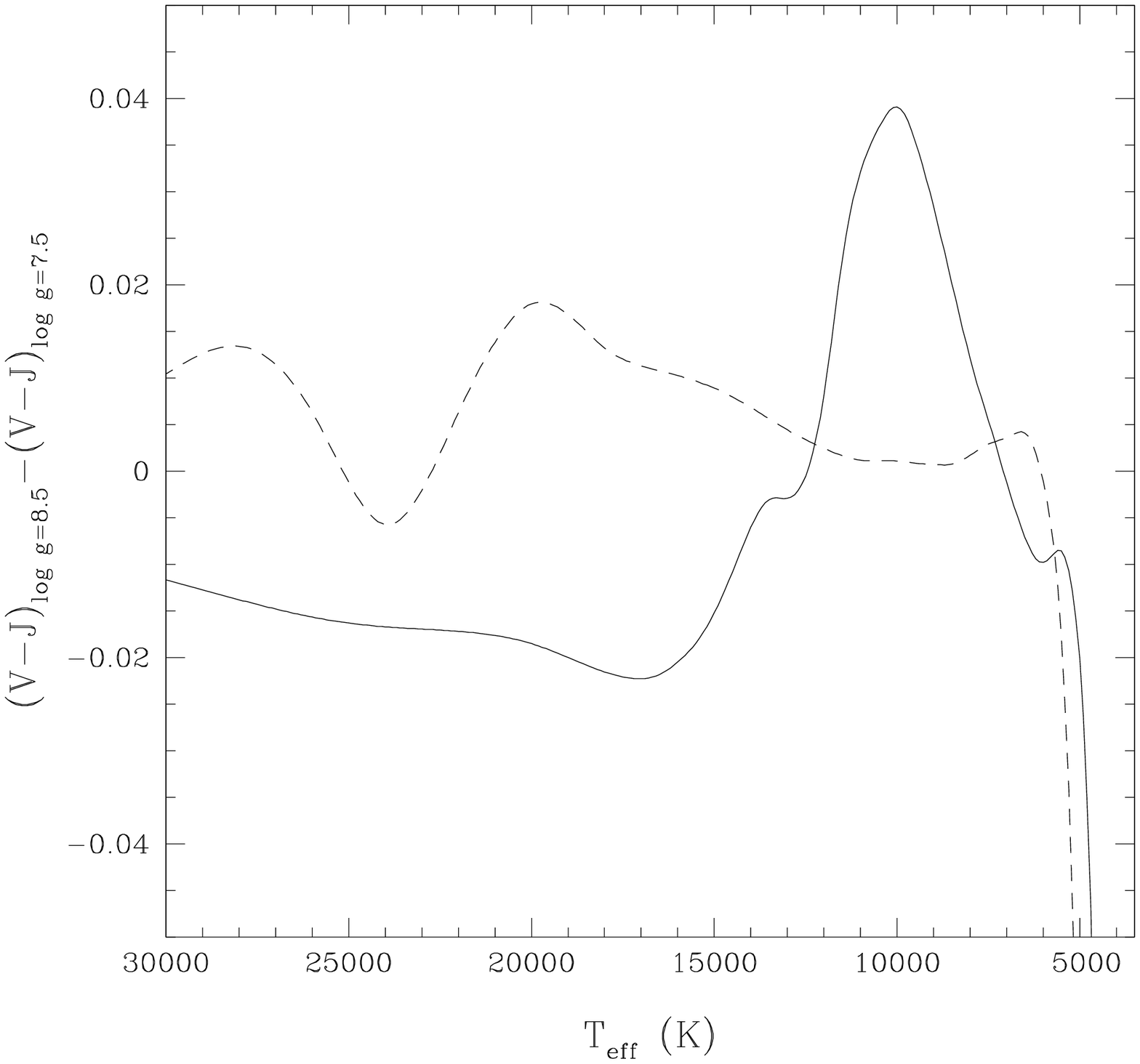]
{Difference in $(V-J)$ color indices as a function of $\Te$ between
$\log g=8.5$ and 7.5 model atmospheres with pure hydrogen ({\it solid
line}) and pure helium ({\it dashed line}) compositions. The general
behavior of these curves can be explained in terms of the various
opacity sources that peak at slightly different effective temperatures
for these two $\log g$ values.\label{fg:f4}}

\figcaption[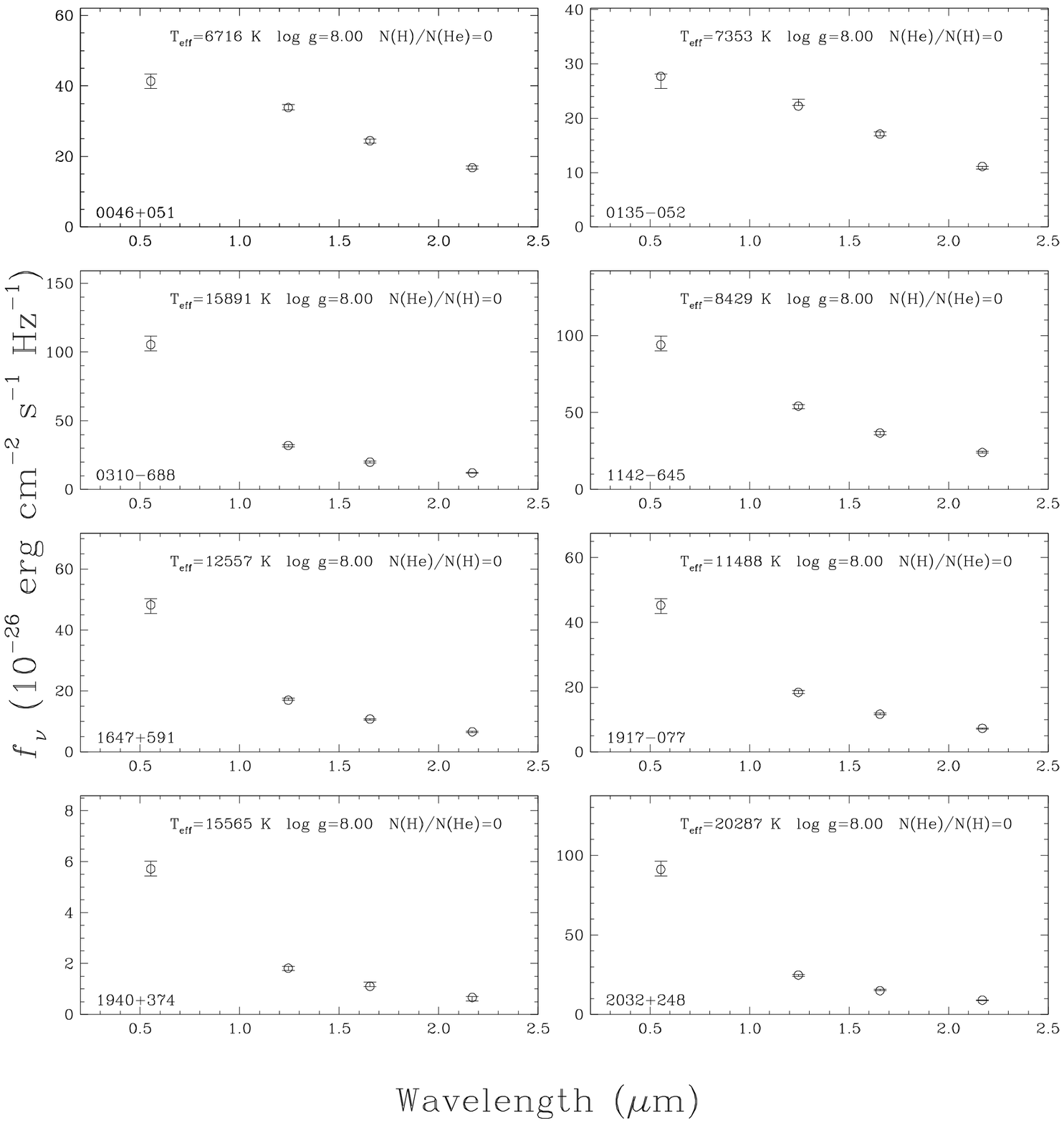]
{Sample fits to the observed $VJHK_S$ photometric observations ({\it
error bars}) of eight bright white dwarfs with hydrogen and helium
atmospheres. The model fluxes are shown by open circles and the
atmospheric parameters are given in each panel.\label{fg:f5}}

\figcaption[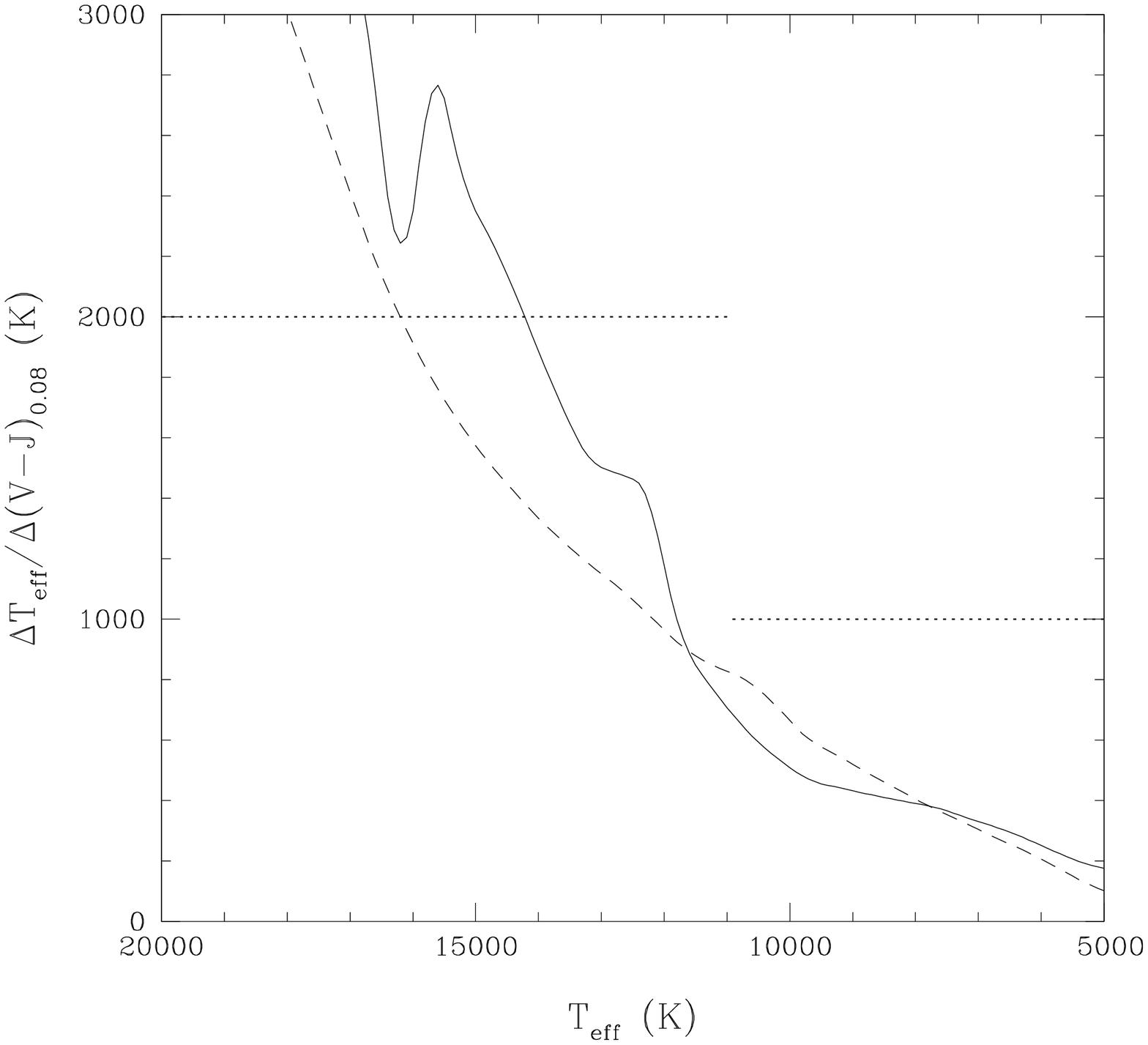]
{Effective temperature variation, $\Delta \Te$, for a corresponding
change of the $(V-J)$ color index of $\Delta (V-J)=0.08$ as a function
of $\Te$ for our pure hydrogen ({\it solid line}) and pure helium
({\it dashed line}) model atmospheres. The dotted horizontal lines
represent the various bin widths of the histograms used in
\S~5.\label{fg:f6}}

\figcaption[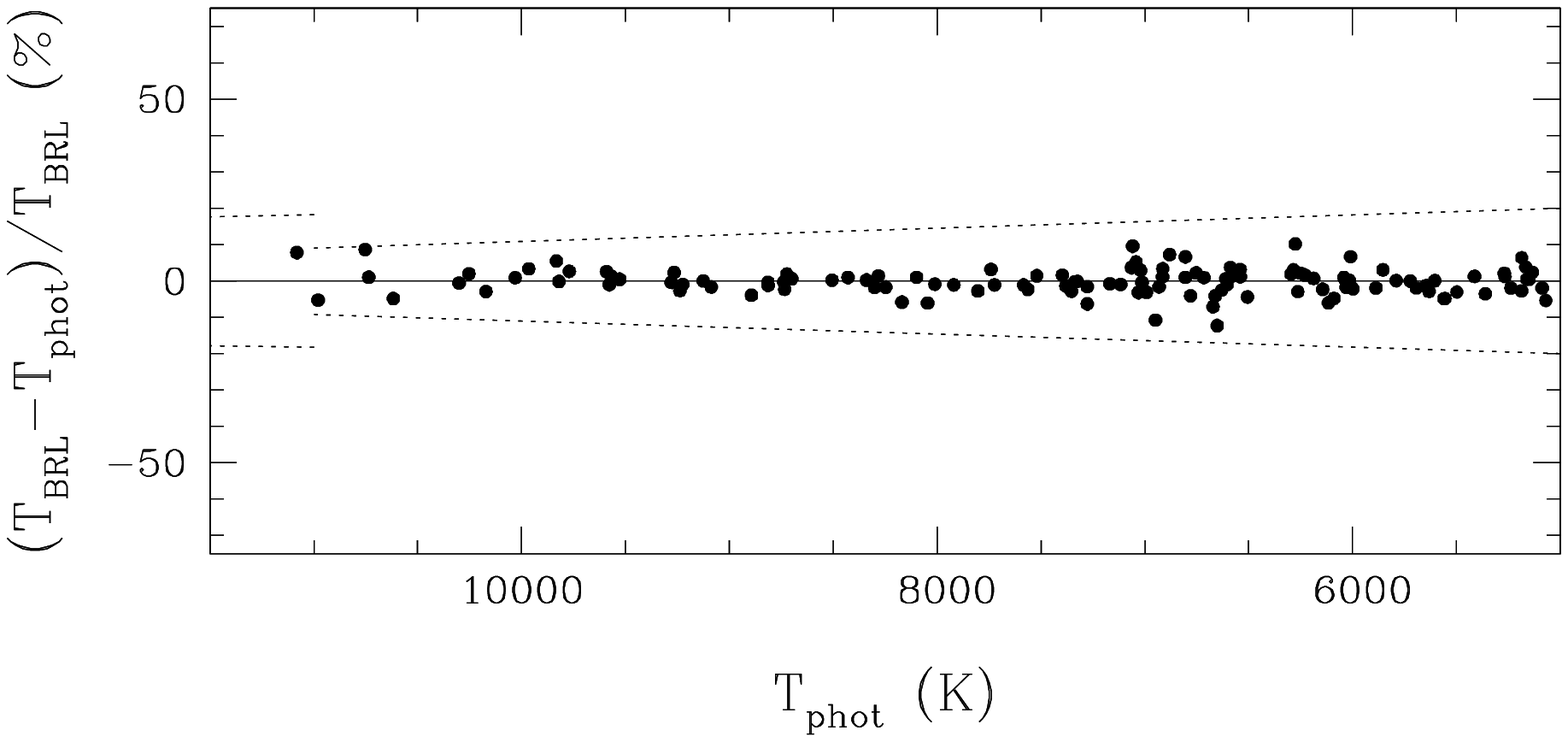]
{Differences in effective temperature determinations based on our
2MASS sample ($T_{\rm phot}$) and the photometric $BVRIJHK$ data
($T_{\rm BRL}$) of \citet{bergeron97,bergeron01} for the 133 cool
white dwarfs in common. The solid line represents a perfect match
between both data sets, while the dashed lines correspond to
temperature uncertainties equal to the bin widths of the histograms
used in \S~5.\label{fg:f7}}

\figcaption[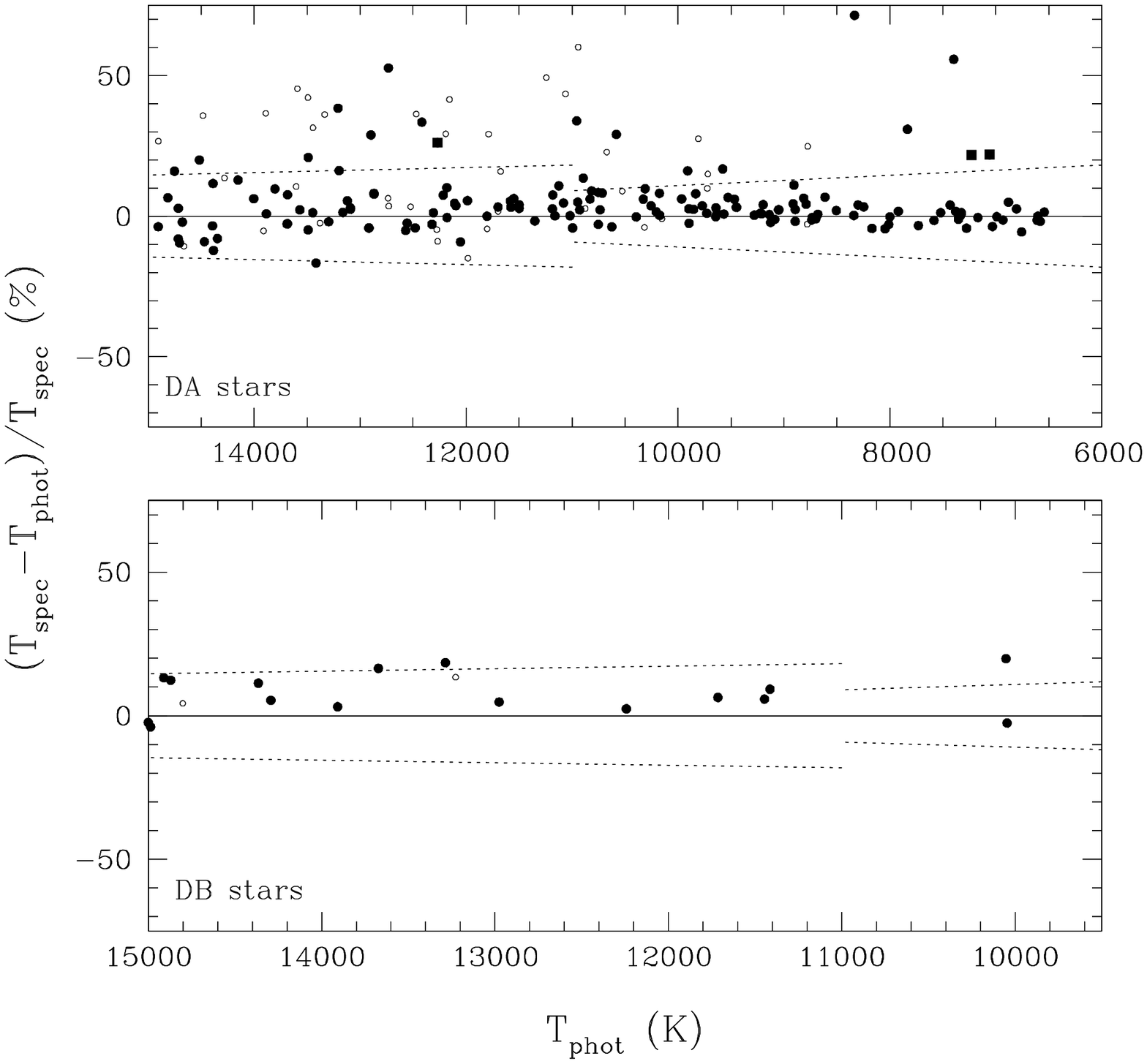]
{\textit{Top panel:} Differences between our photometric $\Te$ values
and those determined spectroscopically for a common sample of 198 DA
white dwarfs. The small open circles represent stars for which the
2MASS fluxes are close to the detection limit with formal
uncertainties in only two bands. The filled squares correspond to
known double degenerate systems. The horizontal and dashed lines have
the same meaning as in Figure \ref{fg:f7}. \textit{Bottom panel:} Same
as top panel but for 18 DB white dwarfs.
\label{fg:f8}}

\figcaption[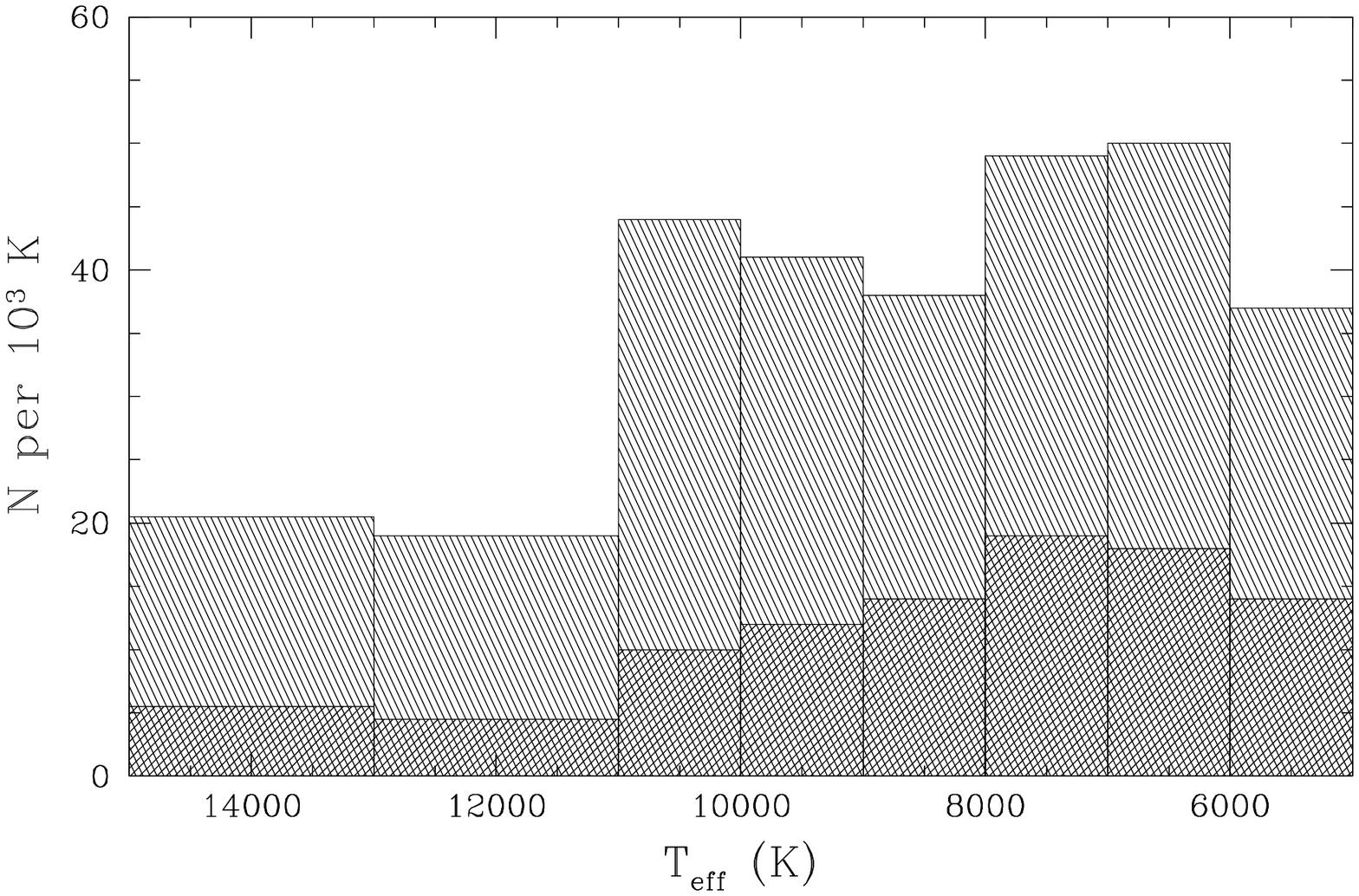]
{Histogram of the number of hydrogen-atmosphere ({\it single hatched})
and helium-atmosphere ({\it double hatched}) white dwarfs from the
2MASS sample as a function of $\Te$. The number of stars in each bin
is per unit of $10^3$~K in $\Te$.
\label{fg:f9}}

\figcaption[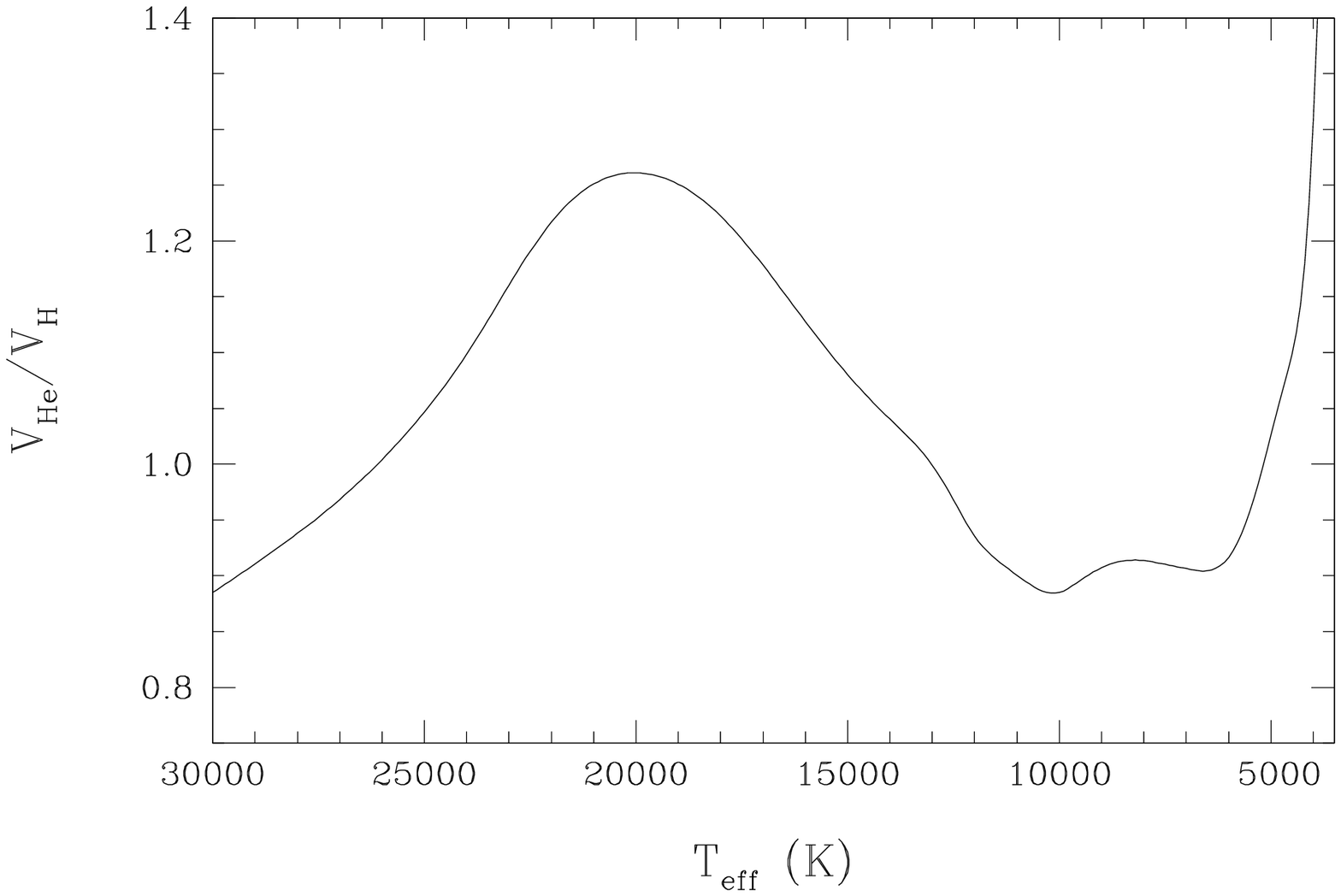]
{The ratio of the observed space volume for helium-atmosphere and
hydrogen-atmosphere white dwarfs in a $H$ magnitude limited survey as a
function of $\Te$.
\label{fg:f10}}

\figcaption[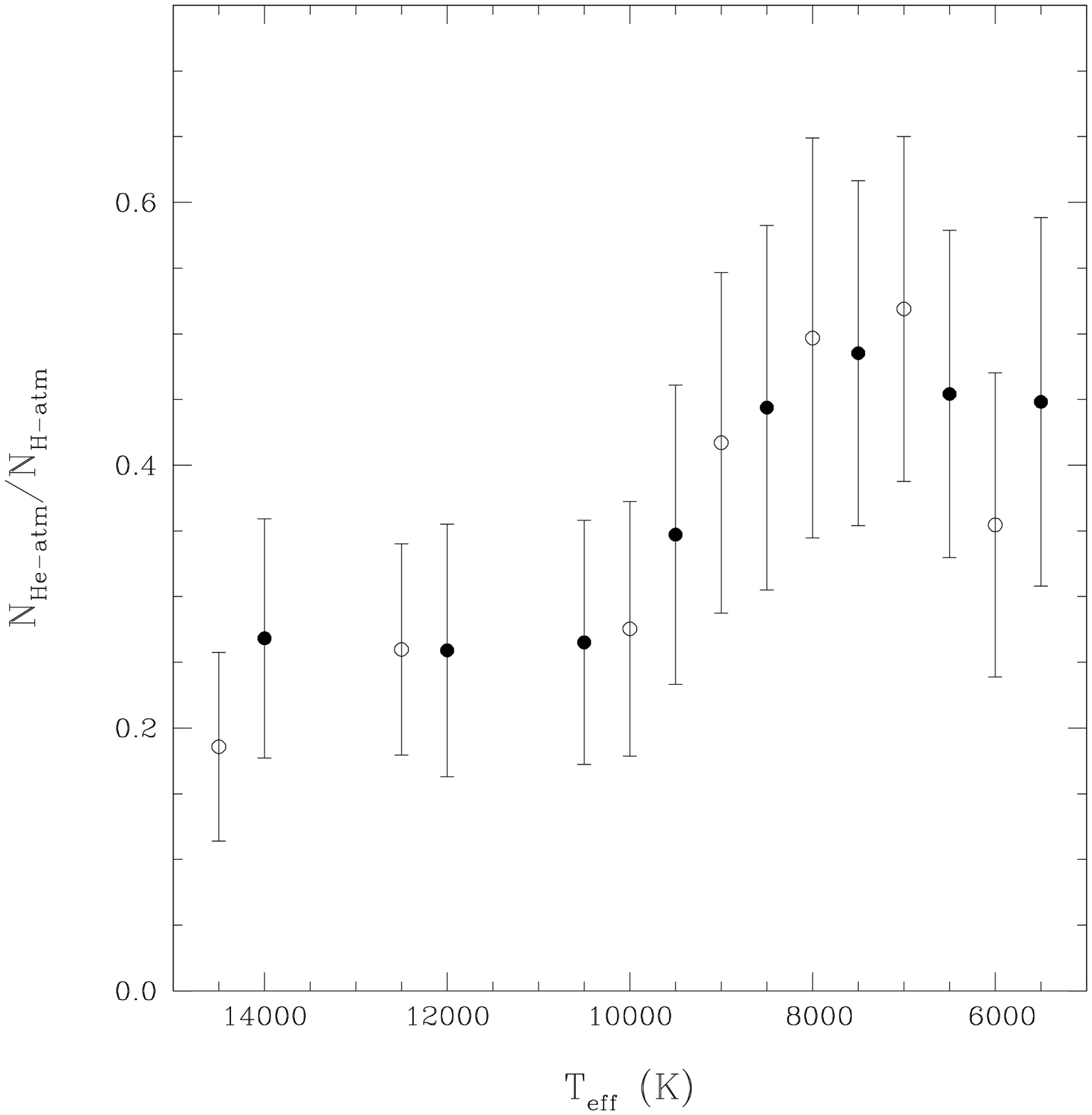]
{The ratio of the number of helium- to hydrogen-atmosphere white dwarfs 
as a function of effective temperature, corrected for the volume
effect displayed in Figure \ref{fg:f10} and discussed in the
text. The filled circles use the binning defined in Figure
\ref{fg:f9} while the open circles present the results
for the temperature bins shifted by 500~K. The error bars are
statistical.
\label{fg:f11}}

\begin{figure}[p]
\plotone{f1.eps}
\begin{flushright}
Figure \ref{fg:f1}
\end{flushright}
\end{figure}

\begin{figure}[p]
\plotone{f2.eps}
\begin{flushright}
Figure \ref{fg:f2}
\end{flushright}
\end{figure}

\begin{figure}[p]
\plotone{f3.eps}
\begin{flushright}
Figure \ref{fg:f3}
\end{flushright}
\end{figure}

\begin{figure}[p]
\plotone{f4.eps}
\begin{flushright}
Figure \ref{fg:f4}
\end{flushright}
\end{figure}

\begin{figure}[p]
\plotone{f5.eps}
\begin{flushright}
Figure \ref{fg:f5}
\end{flushright}
\end{figure}

\begin{figure}[p]
\plotone{f6.eps}
\begin{flushright}
Figure \ref{fg:f6}
\end{flushright}
\end{figure}

\begin{figure}[p]
\plotone{f7.eps}
\begin{flushright}
Figure \ref{fg:f7}
\end{flushright}
\end{figure}

\begin{figure}[p]
\plotone{f8.eps}
\begin{flushright}
Figure \ref{fg:f8}
\end{flushright}
\end{figure}

\begin{figure}[p]
\plotone{f9.eps}
\begin{flushright}
Figure \ref{fg:f9}
\end{flushright}
\end{figure}

\begin{figure}[p]
\plotone{f10.eps}
\begin{flushright}
Figure \ref{fg:f10}
\end{flushright}
\end{figure}

\begin{figure}[p]
\plotone{f11.eps}
\begin{flushright}
Figure \ref{fg:f11}
\end{flushright}
\end{figure}

\end{document}